# Project CERES: Nuclear Thermal Transportation and In-Situ Propellant Production

## A Foundational Architecture for Solar System Logistics

S. P. Worden and F.G. Kennedy

# Abstract

Project CERES proposes a reusable deep-space transportation architecture centered on the dwarf planet Ceres, combining nuclear thermal propulsion (NTP) with in-situ production of water, hydrogen, oxygen and ammonia. Its first flight, a Pathfinder that surveys Ceres and returns samples, also bears on the question of life: Dawn found the ingredients of prebiotic chemistry there, and returned material would show just how far that chemistry went. We survey every direct Earth-Ceres opportunity from 2032 through 2042 with ephemeris-based Lambert analysis and compare chemical, gravity-assisted, nuclear-electric and nuclear thermal vehicles under one parametric mass model. Impulsive requirements vary from 10.2 to 14.1 km/s across the decade. A 900 s NTP stage delivering 35 t closes near 220 t in low Earth orbit and does so at every surveyed window; chemical delivery is marginal and sensitive to opportunity and staging, and nuclear-electric propulsion requires multi-megawatt power to match one-to-two-year trip times. Refueling at Ceres reduces the launch mass of an Earth-Ceres-Earth sortie by 61%, and the specific impulse of the propellant the node produces decides whether Ceres water becomes cargo or tanker mass. We conclude that Ceres uniquely combines abundant volatiles, exceptionally low gravity and a strategic Main Belt position, and that nuclear thermal propulsion is the only propulsion class that closes mass, time and the return leg together. Together they convert deep-space exploration from independent expeditions into a network operating on a timetable.

| Property | Value | Property | Value |
|---|---|---|---|
| Semi-major axis | 2.767 AU | Mean radius | 469.7 km |
| Eccentricity | 0.079 | Mass | $9.38 \times 10^{20}$ kg (~1/3 of belt total) |
| Inclination | 10.59° | GM | 62.63 $km^3/s^2$ |
| Orbital period | 4.60 yr | Surface gravity | 0.28 $m/s^2$ (0.028 g) |
| Earth synodic period | 1.28 yr (15.3 mo) | Escape velocity | 0.51 km/s |
| Rotation period | 9.07 h | Low-orbit (100 km) speed | 332 m/s |
| Solar flux | 13% of Earth's | Surface temperature | ~110–155 K |
| Discovered | 1801 (Piazzi) | Explored | Dawn orbiter, 2015–2018 |

*Table 1. Ceres vital statistics.*

# 1. Ceres as port of entry

On 1 January 1801 Giuseppe Piazzi found Ceres where the Titius-Bode progression had predicted a planet; when it was lost behind the Sun weeks later, Carl Friedrich Gauss invented the method of orbit determination that recovered it [1]. Within a few years Pallas, Juno and Vesta showed that the missing planet was a belt. Ceres has since been classified as a planet, an asteroid and a dwarf planet, and the object that motivated modern celestial mechanics is now the object whose accessibility we examine.

Ceres (Fig. 1, Table 1) formed from silicates, water ice and volatile-rich compounds, and heat from radioactive decay drove a global subsurface ocean that left hydrated silicates, carbonates, ammoniated minerals and organics before it froze into an ice-rich crust that may still overlie brines. NASA's Dawn orbiter (2015–2018) measured hydrogen consistent with water ice near the surface at mid-to-high latitudes, mapped hydrated minerals and sodium carbonates, found ammoniated phyllosilicates distributed across the surface and localized aliphatic organics, and recorded geologically recent cryovolcanism at Ahuna Mons and brine deposits in Occator Crater [2, 3]. Liquid water, organic carbon, nitrogen and an internal heat source are the ingredients of prebiotic chemistry, and the 2023 planetary science decadal survey identified Ceres sample return as a priority on those grounds [3]. Whether Ceres went further than chemistry is a question only returned material can settle.For a transportation architecture the inventory is water for direct reaction mass, hydrogen and hydrolox; nitrogen-bearing minerals for ammonia and industrial chemistry; and carbon in carbonates and organics. Unlike almost any other small body, these occur together on a single world that has already been explored in detail.

For more than sixty years every deep-space mission has been architecturally self-contained: each spacecraft carried its own transportation system, launched all of its propellant from Earth's surface, and left no reusable capability behind. Terrestrial exploration became sustained commerce only when infrastructure replaced expeditions. A permanent node beyond Earth requires abundant extractable resources, a gravity well shallow enough to move them economically, and a location from which onward travel is energetically favorable. Among the bodies presently characterized, Ceres combines these more completely than any other, and its 15-month synodic period with Earth supports campaigns rather than isolated events.

Three features carry the case. Its volatiles support three propellant chains in ascending order of processing burden: water heated directly in a nuclear thermal engine (specific impulse near 390 s, a melt-and-filter step) [4]; ammonia (about 500 s, a storable soft cryogen); and electrolytic hydrogen (900 s, requiring electrolysis, liquefaction and zero-boiloff storage), so the node never depends on a single product. Its gravity well is close to a formality: escape velocity is 0.51 km/s, surface gravity 0.028 g, low-orbit speed 332 m/s, and a round-trip tanker sortie costs under 1,000 m/s at a mass ratio of 1.3, so landing a hundred-tonne vehicle resembles a docking maneuver more than a landing. And it sits at 2.77 AU (Fig. 2), two-thirds of the way out of the Sun's gravity well in energy terms: departure hyperbolic excess to Jupiter is 2.6 km/s from Ceres against 8.8 km/s from Earth, and solar-system escape requires 7.4 km/s against 12.3.

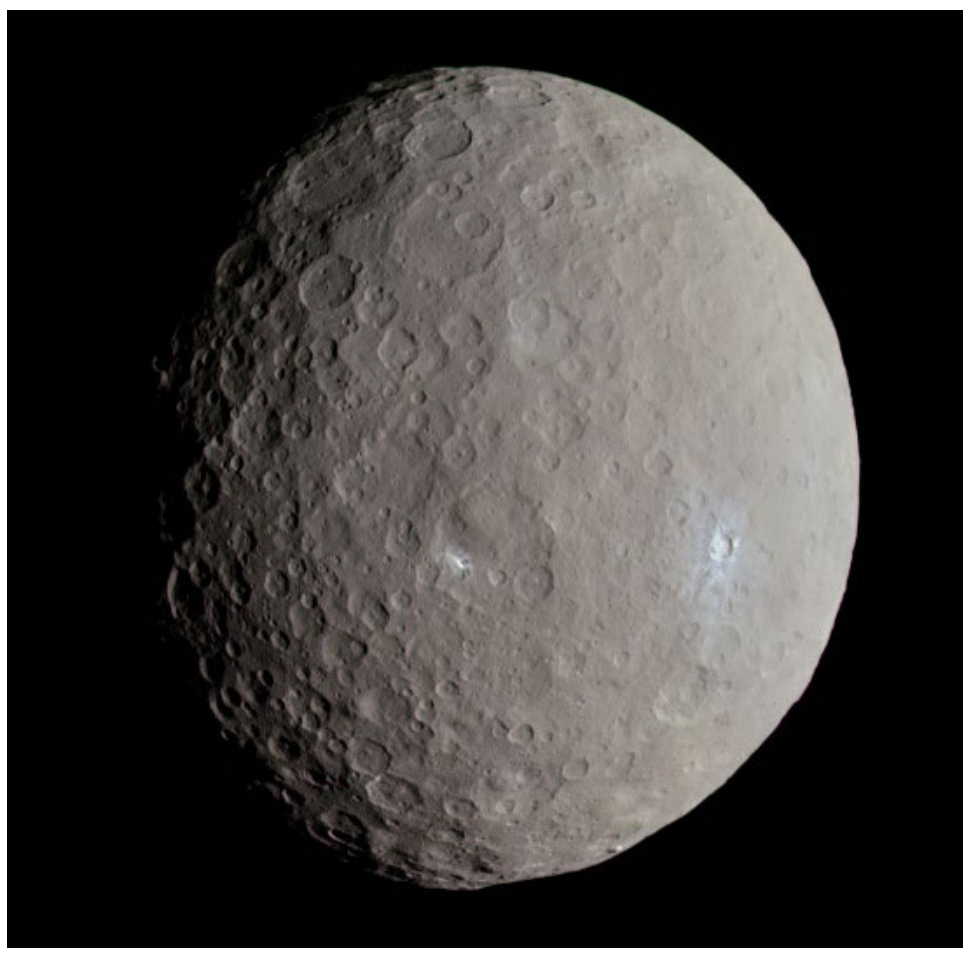

*Figure 1. 1 Ceres, photographed by NASA's Dawn mission in 2015 (real-color).*

Project CERES (Ceres-Earth Resource Exchange System) is the proposal that follows: a reusable logistics network centered on Ceres, combining nuclear thermal propulsion [5] with in-situ use of its volatiles, that replaces the isolated expedition with reusable vehicles, local propellant and permanent infrastructure, for science at scale and tempo, for industrialization, and for connecting the outer planets and their moons with Earth, the Moon and Mars.

## 2. The transportation problem

Direct transits to Ceres are expensive; Mars is a poor guide. Ballistic Earth-Ceres transfer demands a departure characteristic energy (C3) of 43–45 km²/s² at the best opportunities of the coming two decades, roughly four times that of a favorable Mars window, and Ceres offers neither an atmosphere for aerocapture nor a gravity well deep enough to assist propulsive capture: an arrival hyperbolic excess of 5–6 km/s must be removed almost entirely by the propulsion system, against an orbital velocity at low altitude of only 330 m/s. Summed from a 400 km Earth parking orbit to a 100 km Ceres orbit, the one-way impulsive requirement at the most favorable windows of the 2030s is 10.2–11.2 km/s. That exceeds the total ΔV of most Earth-Mars round trips.

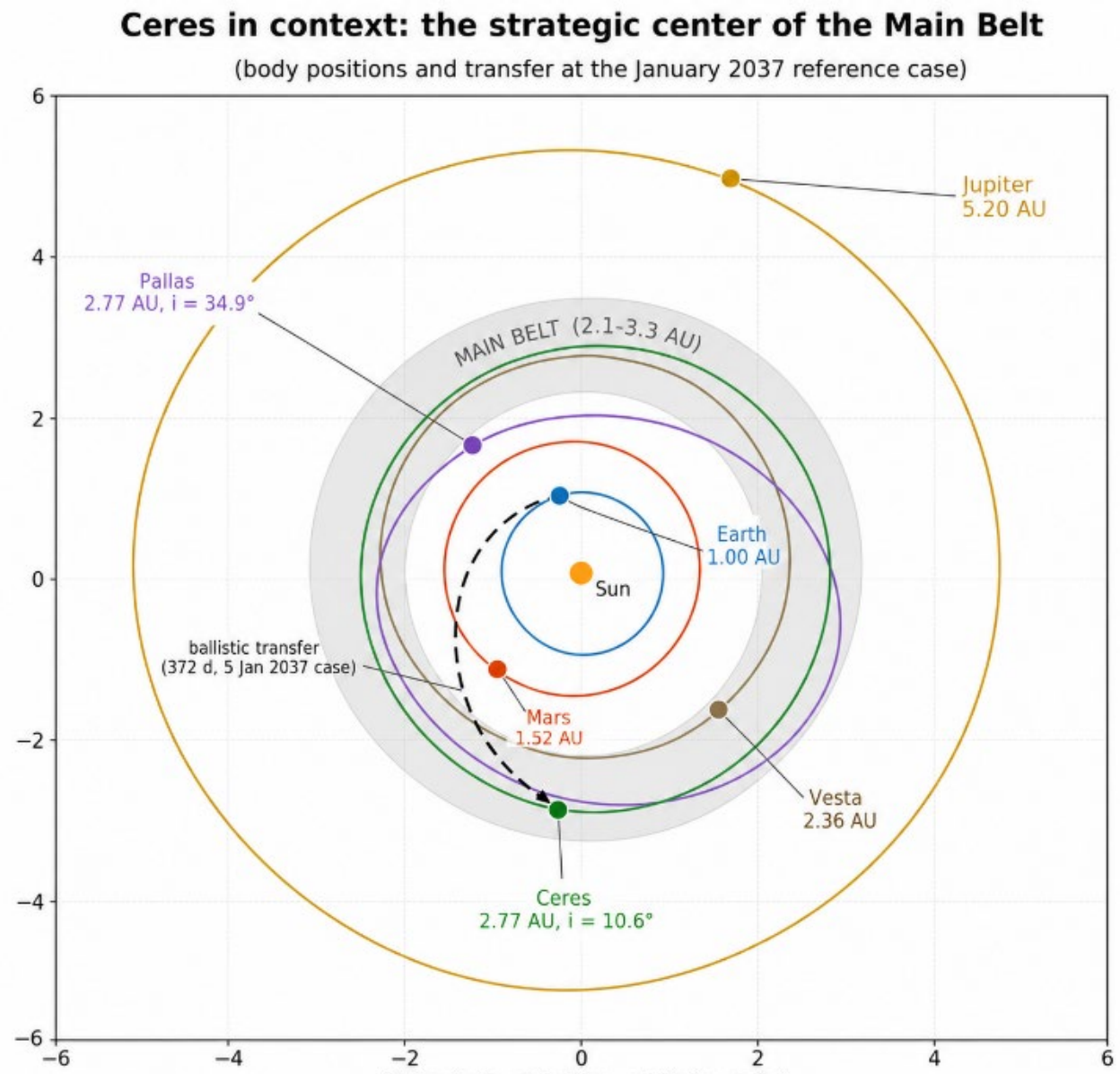

*Figure 2. Ceres in context (orbits to scale, ecliptic projection). Ceres orbits at 2.77 AU in the heart of the Main Belt; a ballistic Earth transfer is shown dashed. Pallas, nearly matching Ceres in heliocentric distance, is inclined 34.9°.*

Ceres' orbit is inclined 10.6° to the ecliptic, so the quality of each opportunity depends on where its transfer geometry falls relative to the line of nodes. Our survey of every ballistic opportunity from 2032 to 2042, computed against JPL Horizons ephemeris states rather than mean elements, shows per-window optima ranging from 10.2 to 14.1 km/s (Fig. 3): a spread of nearly 4 km/s within a single decade, with favorable geometry recurring only every third or fourth synodic period. Published studies corroborate the spread indirectly through plane-change and capture penalties and through strong dependence on launch epoch [6]; we found no published decadal survey of ballistic Earth-Ceres accessibility.

Two methodological notes: (1) Ceres states must come from an ephemeris - propagating catalog J2000 mean elements accumulates along-track errors of up to 27% of an AU and 1.8 km/s over the 2030s, enough to displace computed windows by weeks and hundreds of m/s; and (2) the propulsion comparisons that follow use a single parametric stage mass model, applied identically to every propulsion class.

# 3. The propulsion trade

Can any propulsion type offer frequent, repeatable visits to Ceres given these constraints? Three are candidates. We will test each against a requirement to transport 35 t from low Earth orbit to low Ceres orbit.

*Chemical propulsion.* When hydrogen-oxygen stages at 450 s are burdened with realistic tankage and structure, deliverable payload reaches zero at a stage ΔV of 12.0 km/s under the single-stage mass model, regardless of scale: no quantity of propellant aggregation rescues a stage whose tanks grow as fast as its propellant load. Five-percent payload delivery requires a one-way ΔV below 9.7 km/s, which no direct opportunity in the surveyed decade provides. Single-stage delivery exceeds one percent at only the two best windows of the decade, 3.4% at roughly 1,000 t in LEO in June 2034 and about 2% at 1,900 t in January 2037 (Fig. 4). A two-stage, fully expendable stack closes at 3–7% from 500 to over 1,000 t in LEO per 35 t delivered, with every element discarded and 12 to 26 months of cryogenic propellant retention left uncharged; charged with space-storable propellants at 320 s, it does not close at all.

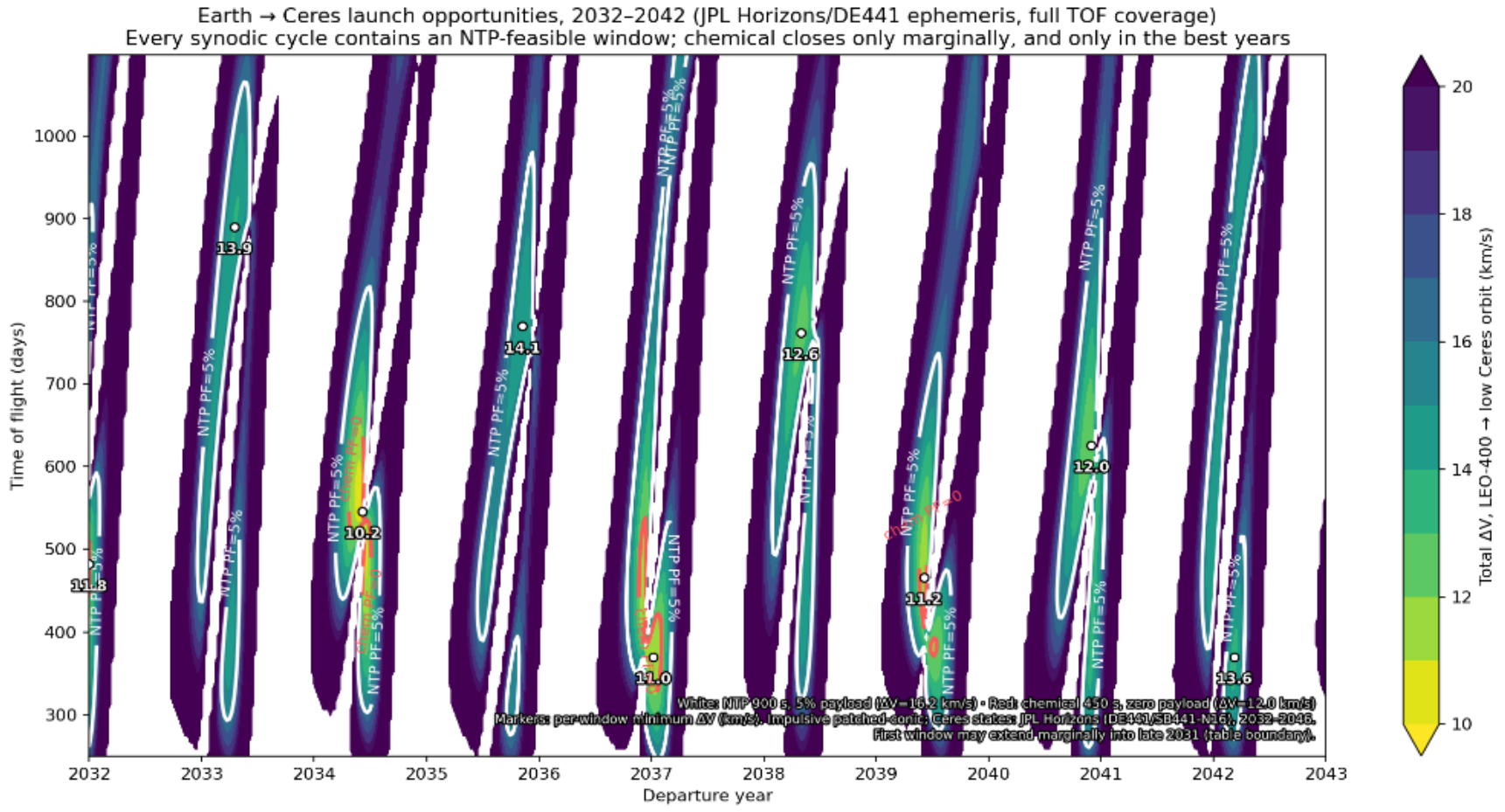

*Figure 3. Earth → Ceres launch opportunities, 2032–2042 (JPL Horizons/DE441 ephemeris; impulsive patched-conic). White contour: NTP 5%-payload feasibility ceiling (16.2 km/s); red: chemical single-stage zero-payload ceiling (12.0 km/s). Markers: per-window minimum ΔV. Every synodic cycle contains an NTP-feasible window.*

An established remedy for the direct route's cost is a Mars gravity assist [7]. We surveyed the Earth-Mars-Ceres pathway with the same toolchain and deliberately generous assumptions. At its best alignment it delivers the mission for 9.4 km/s, below the best direct opportunity, and a 450 s stage then places 5–6% of a 630 t stack at Ceres. That alignment occurs precisely once in the decade, a three-week period in September 2037 - and the flight takes three years. Two further clusters, in mid-2035 and mid-2040, sit at about 11.6 km/s and deliver at the 1–2% level; there is nothing else. Multi-revolution direct transfers tell the same story in miniature: a one-revolution trajectory improves the worst direct window from 14.1 to 12.6 km/s, at flight times beyond 3.3 years. Every workaround that restores chemical mass closure, the flyby, the phasing leg, the extra revolution, pays for it in time and schedule.

*Electric propulsion.* Solar-electric spacecraft have rendezvoused with both Vesta and Ceres. The low-thrust ΔV for the same delivery, including spiral escape, the 10.6° plane change and capture spiral, is 21.6 km/s. Nuclear-electric propulsion closes the mass equation handsomely; time is the difficulty. Matching the nuclear-thermal strong-window trip times of 1.0 to 1.3 years requires about 1.5 MWe under the most optimistic published specific-power assumptions and 2–3 MWe under flown-class ones [8]. The most powerful space nuclear power reactor flown to date produced about 6 kWe. At the tens to hundreds of kilowatts a near-term program could credibly field, one-way trip times run four to ten years: Dawn's own arithmetic, and Dawn was not keeping a schedule. We endorse electric propulsion for what it is suited to, time-insensitive bulk cargo delivered to a network whose timetable is held by high-thrust vehicles, and expect the two classes to coexist.

| **Propulsion class** | **Payload fraction, best window (impulsive)** | **LEO mass per 35 t delivered** | **One-way time** | **Windows ≥5% per decade** | **Return-leg mass ratio on Ceres propellant** | **Reusable on a schedule** |
|---|---|---|---|---|---|---|
| Chemical, single stage (450 s) | 3.4% (Jun 2034) | ~1,000 t | 1.0–1.3 yr | 0 | 4.0 (390 s) | No |
| Chemical, two stage, expendable | 3–7% (Jun 2034) | 500–1,000+ t | 1.0–1.3 yr | 2 (marginal) | 4.0 | No |
| Chemical via Mars assist | 5–6% (Sep 2037) | ~630 t | 3.0 yr | 1 | 4.0 | No |
| Nuclear electric, 100s of kWe | closes | small | 4–10 yr | n/a | n/a | No (cadence) |
| Nuclear electric, ≥1.5 MWe | closes | small | 1.0–1.3 yr | n/a | n/a | Requires ~100× flown reactor power |
| Nuclear thermal, 900 s | 19.9% (Jun 2034); 17.4% (Jan 2037) | 176–201 t impulsive; 220 t with losses | 1.0 yr | 9 (every cycle) | 1.82 | Yes |

*Table 2. Propulsion-class comparison for a 35 t delivery from low Earth orbit to low Ceres orbit, 2032–2042. All values from the Section 2 survey and the single parametric mass model; chemical cases are given best geometry, staging, and no boiloff charge.*

*Nuclear thermal propulsion.* At 900 s, the June 2034 route peaks near 19.9% payload fraction (Fig. 4); the January 2037 reference case of Section 3.1 yields 17.4% at a 372-day transit, 16% with finite-burn losses (the 220 t closure of Section 3.1). NTP remains above 14% across a broad range of flight times and retains 5% capability out to 16.2 km/s, which covers every window in the surveyed decade (Fig. 3). Solid-core NTP is the only class that pairs chemical-class thrust with roughly double chemical specific impulse, and that pairing is what holds a timetable across the geometries the solar system actually presents.

Table 2 is the comparison. The last two columns are the ones this paper turns on. A vehicle that is to fly again must return to cislunar space on propellant made at Ceres, and the mass ratio for that return leg depends on the specific impulse of the propellant the node produces: 1.82 at 900 s, 4.0 at 390 s. Whether Ceres water becomes cargo or becomes tanker mass is decided by the engine.

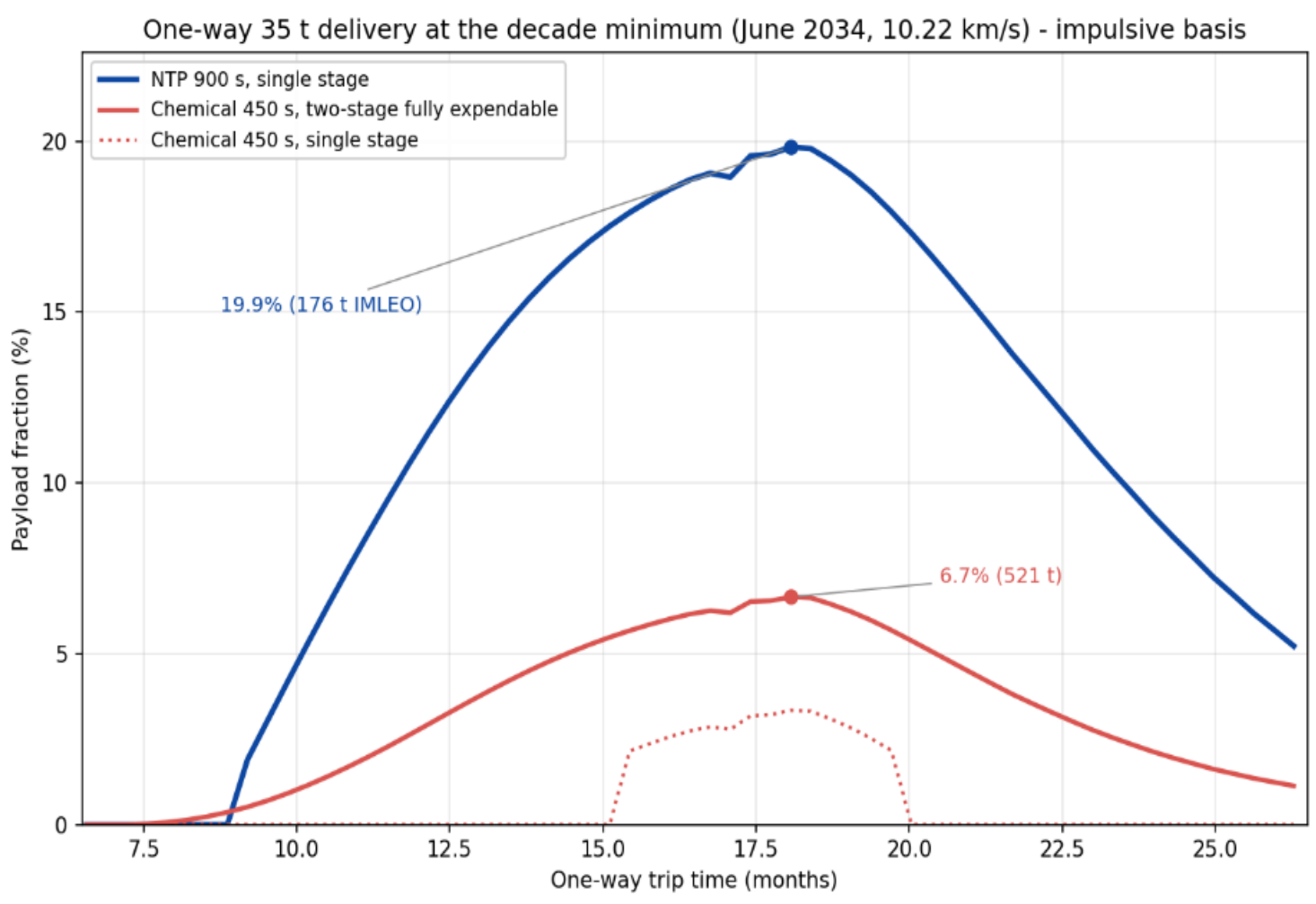

*Figure 4. One-way delivery, 35 t payload, at the decade's best window (Jun 2034): payload fraction vs trip time, with chemical given best geometry, staging, and no boiloff charge.*

The objection this table answers is a reasonable one. Chemical propulsion has flown to every planet, and a chemical Ceres mission is within the reach of any capable stage builder. That is true, and it is a different question. A mission is one flight at the best window of a decade. A logistics node is a vehicle that flies every window, returns, and flies again. Chemical propulsion can deliver a few percent of its launch mass to Ceres at two opportunities in ten years, cannot bring the vehicle home on Ceres propellant without a mass ratio of four, and cannot hold a timetable. Electric propulsion holds the mass equation and loses the calendar. Nuclear thermal propulsion closes mass, time and the return leg together, at every synodic opportunity surveyed. Ceres missions are possible without it. A Ceres node is not.

Chemical stages retain a role in the architecture at the Earth end, in launch and cislunar assembly, where their thrust is an asset and their specific impulse is not the constraint.

## 3.1 A Representative Vehicle

The reference vehicle (Fig. 5) for this study delivers 35 t of payload from low Earth orbit to low Ceres orbit and is sized by the January 2037 selected reference case (10.95 km/s impulsive one-way; chosen for Pathfinder scheduling - the decade minimum is June 2034 at 10.22, Fig. 3) with finite-burn losses applied (departure 6.9%, capture 1.4%; derivation in the supplementary package).

At 900 s specific impulse the vehicle closes at 220 t gross mass in LEO for a 372-day transit (design-envelope sizing at C3 = 44.5 $km^2/s^2$, covering every window of the decade; the specific January 2037 closure is 218.5 t). The propulsion system is a cluster of six solid-core engines in the 6,000-lbf (27 kN) thrust class, each with reactor-and-engine mass of approximately 500 kg - a thrust class consistent with particle-bed reactor designs offering engine thrust-to-weight between 5 and 15.

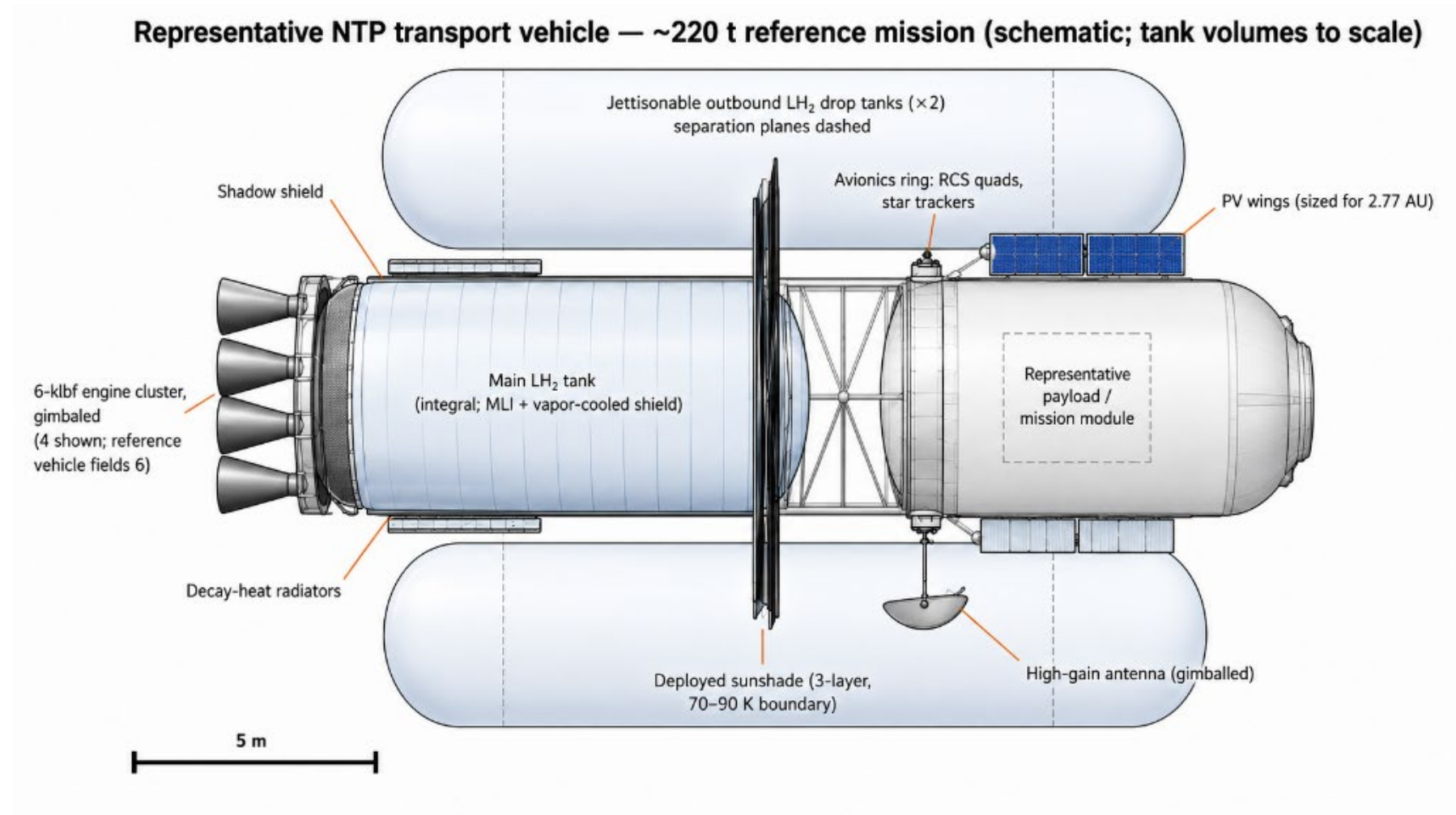

*Figure 5. Representative 220 t orbital transfer vehicle for Earth-Ceres-Earth missions. Six 6 klbf NTP engines, $LH_2$ propellant (Isp = 900 s).*

Clustering is a design position: it provides engine-out capability on a nuclear stage (a single-engine failure extends burn schedules rather than ending the mission), permits engine-level ground qualification at manageable test-facility scale, and holds a single engine design across a payload range from 20 t (three engines) to 50 t and beyond (eight engines), preserving fleet commonality.

At 220 t gross, six engines yield an initial thrust-to-weight of 0.074.

Restart count remains within the demonstrated envelope of Rover/NERVA ground testing.[9] Ceres capture splits the arrival burn: most of the hyperbolic excess is removed in the far field, where Ceres' weak gravity makes finite-burn loss small, and a periapsis burn of under ten minutes retains the limited Oberth benefit. The integrated penalty is 1.4% over the impulsive value for the reference configuration; 1.8% is retained in the mass audits as a conservative bound.

Note: 220 t is this vehicle's one-way mass in LEO, and because the return leg is fueled at Ceres, it is also its round-trip mass. Carry that return propellant from Earth instead and the same trip launches 338 t heavier: nearly seven kilograms lifted off Earth for every kilogram burned on the way home.

# 4. No new physics

The case of Section 3 requires no new physical principle. Solid-core NTP (Fig. 6) has an unusually substantial ground-test heritage among advanced propulsion concepts: NASA's Rover/NERVA conducted more than twenty reactor and engine tests, demonstrated fuel temperatures consistent with a specific impulse near 850 s, achieved repeated restart, and operated across thrust levels from roughly 25 klbf to more than 200 klbf [9]. The 900 s assumed in this study is a development objective, not a demonstrated flight value. Reaching it requires higher-temperature fuel, qualification, and integrated stage development beyond the NERVA state of the art. This qualification is a Phase 1 deliverable: it is a materials program, not a physics one - a modest extrapolation from ground-tested hardware against which the several-hundred-fold power extrapolation of the nuclear electric alternative should be read.

Two later efforts define the contemporary design space. The Space Nuclear Thermal Propulsion program (1987–1994) pursued the particle-bed reactor, with sub-millimeter coated fuel particles retained in annular beds and cooled by radially inflowing hydrogen; its projected advantages were compactness, high power density and engine thrust-to-weight well above the NERVA-derived baseline, and those values remain design projections [10]. The NASA/DARPA DRACO partnership advanced reactor and integrated-stage design before closing in April 2025 without a flight demonstration, and subsequent federal direction consolidated behind fission power and nuclear electric propulsion, leaving nuclear thermal propulsion without a government flight program [11]. The vehicle of Section 3.1 assumes an SNTP-derivative engine. Whether the capability is fielded in the coming decade is therefore a question of program, not of physics; this paper's purpose is to show what the program would buy.

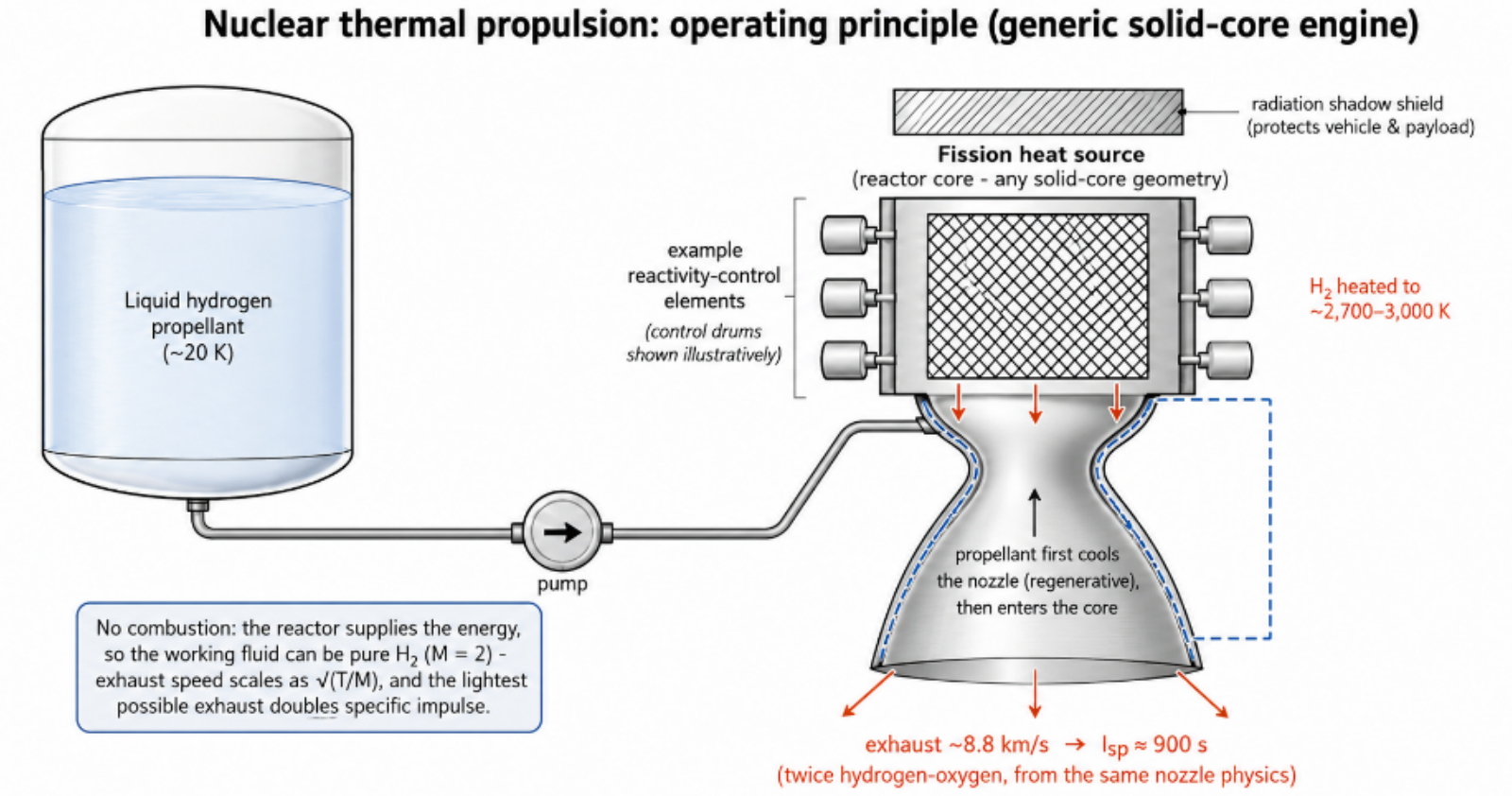

*Figure 6. Nuclear thermal propulsion, operating principle (generic solid-core engine; no specific core geometry implied). Illustrative reactivity-control elements are shown as one representative approach. 900 s specific impulse, $LH_2$ propellant.*

# 5. The node and the return leg

Transportation alone does not create infrastructure. The architecture is a hub-and-spoke logistics network with its hub at Ceres, providing four services in order of increasing ambition: a propellant depot, ferry turnaround and refit for vehicles that never return to Earth's gravity well, staging and assembly for onward legs, and eventually fabrication. Its defining property is the timetable it keeps: departures from Earth every synodic period regardless of window quality, and departures from Ceres whenever physics permits. The unit of analysis is a route rather than a mission, and the test of the architecture is that every route closes with vehicles we can build and propellant we can produce.

The physical layout follows from Ceres' peculiar virtues. Surface gravity of 0.03 g and a low-orbit velocity of 332 m/s make the surface-to-orbit leg nearly free: a round-trip tanker sortie of under 1,000 m/s implies a water-propelled hopper at a mass ratio of 1.3, with propellant about 23% of departure mass. Processing therefore happens on the surface, at an ice-rich high-latitude site selected by the Pathfinder survey, and only finished product goes to orbit. The alternatives are not close: lofting raw regolith to an orbital processor means moving 4,400 t per window at 10 weight percent extractable water; lofting water for orbital electrolysis, 440 t; lofting finished liquid hydrogen, 49 t. The node in low Ceres orbit is a tank farm, a set of berths and a staging volume.

The buildout is a spiral in which each phase delivers standalone value and requires only the previous phase. Phase 0 is the Pathfinder flight of Section 6. Phase 1, the wellhead, is a surface extraction plant and orbital tank farm producing about 180 t of water per synodic window, 390 kg per day, enough to fuel one full vehicle return per cycle on direct steam propulsion at roughly 390 s and a return-leg mass ratio of 4.0 [12]. Extraction at this rate is a 15–18 kWt thermal problem that a modest concentrating solar plant closes even at 13% of terrestrial flux; nuclear power is useful here rather than essential, and the flight reactor should not be credited with surface heat unless the architecture places it there. Phase 1's price of admission is a water-qualified engine variant that accepts a temperature derate and an oxide-tolerant fuel form, since hot steam is aggressive toward carbide fuels.

Phase 2, the refinery, adds electrolysis, liquefaction and zero-boiloff storage, and brings hydrogen online: 49 t of LH2 per window from 440 t of water feed, and return legs at a mass ratio of 1.82. Ore grade enters buildout cost and fleet size linearly, which makes the Pathfinder's grade map the parameter that sizes the excavation plant. The defining resource is power. Electrolysis at 50 kWh per kg of hydrogen and liquefaction at 12 kWh/kg, drawn continuously across the window, come to 274 kWe and 44 kWt, or a few hundred kilowatts electric with parasitics, through Ceres' 9.07-hour nights for years. A photovoltaic plant meeting that duty is in the 40 t class and needs a night store cycled some 970 times a year; a fission surface plant of the same output masses 15–30 t, is indifferent to latitude, and supplies the extraction load and the tank farm's zero-boiloff plant from its waste heat. On mass alone the two are within a factor of two; siting, deployment and the night decide it. This is where the one-nuclear-ecosystem argument becomes concrete: the same industrial and regulatory base that fields the engines fields the power plant.

Phase 3 adds fabrication and the outer-system network of Section 7. Fleet operations are set by the 15.3-month window and a 2.5–3 year round trip, so three vehicles hold a continuous schedule with one outbound, one at the node or returning, and one in cislunar refit. The Phase 2 long-lead package of 30–45 t fits inside a single eight-engine sortie, so the node never requires an on-orbit assembly campaign to stand up its own power source.

The return leg is where the node and the engine meet. A vehicle returning empty to cislunar space draws about 21 t of locally produced hydrogen, or 76 t of water in steam mode; returning loaded with 35 t of product, it draws the full window's allotment of 49 t and 181 t respectively. Every tonne of water that becomes propellant at 390 s rather than 900 s is a tonne the node must mine and lift for nothing.

# 6. The Project CERES Pathfinder Mission

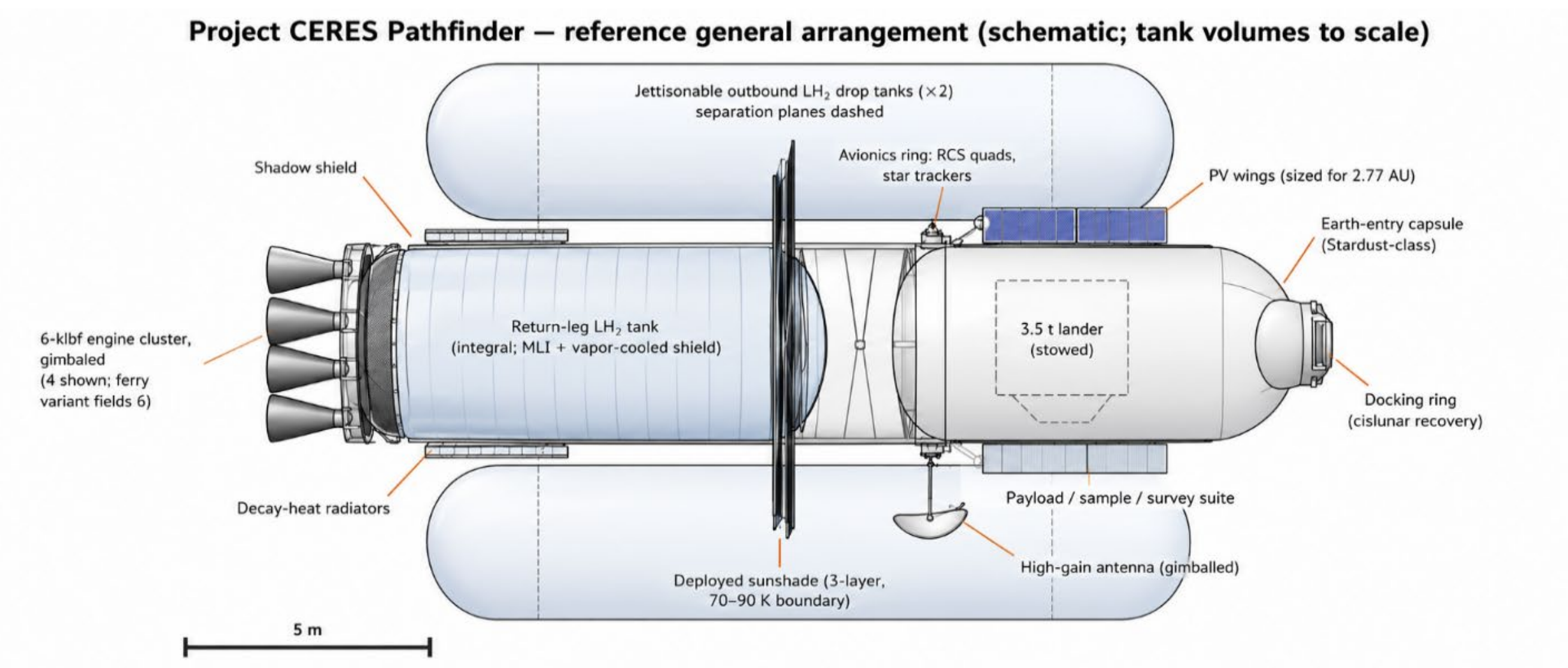

*Figure 7. Project CERES Pathfinder - reference general arrangement (schematic; tank volumes to scale). Four-engine cluster (the ferry variant of Section 3.3 fields six), shadow shield, integral return tank, two jettisonable outbound drop tanks, deployed sunshade, survey payload with stowed lander, Stardust-class entry capsule, and docking ring for cislunar recovery.*

Every architecture in this paper descends from a single flight. Project CERES is that flight: one vehicle, launched once, that crosses the inner solar system in a year, surveys the most volatile-rich world between Earth and Jupiter, touches its surface, extracts water from ground that has held it for four billion years, and comes home. It returns the first samples of a dwarf planet and something rarer than samples: proof that a spacecraft can leave Earth, replenish itself at another world, and take up station in cislunar space intact, fueled by its destination, awaiting its next assignment. Two centuries ago the astronomers of the *Himmelspolizei* organized themselves to find Ceres; we can reach it. The sections preceding this one argue that the architecture closes. The mission described here is where the argument ends and organizing for flight begins.

The samples are also the architecture's first science return: material from a brine-bearing, organic-rich body, chosen after months of orbital survey rather than from a site committed at launch, and in terrestrial laboratories within five years. A conventional sample return could fetch them too, a decade later and from wherever its launch geometry allowed. The Pathfinder's contribution is the choice of site and the calendar.

## 6.1 Flight system

The Pathfinder is a single integrated spacecraft (Fig. 7): a four-engine variant of the Section 3.1 vehicle carrying its own survey payload, a deployable lander for the extraction demonstration, and a Stardust-class return capsule. Every element of the ferry architecture appears here at subscale. Samples return by direct entry from an arrival-optimized June 2040 departure: a 6.9 km/s trans-Earth injection and a 330-day transit deliver the capsule at 13.0–13.1 km/s, the lowest entry speed of any surveyed return window in the decade and close to the 12.9 km/s Stardust record, though proximity to heritage is not qualification and the capsule requires mission-specific analysis and test [13]. It is carried as a containment-qualified element to whatever COSPAR categorization the survey data support [14].

A faster mid-2038 return saves more than a year at 16–19 km/s entry, beyond flown heritage; the baseline trades that year for a lower-risk entry and converts the added Ceres residence into survey operations. The vehicle itself, arriving at about 7 km/s, captures propulsively: a lunar-flyby-assisted periapsis burn of 1.4–1.8 km/s places it in a bound NRHO-class orbit in the month the samples land, closing the return-to-cislunar demonstration in full [15]. The flight vehicle carries no active cryocooling; a deployed sunshade and vapor-cooled shielding hold boiloff to about 2 t across the outbound transit, carried inside the load of Table 3, and zero-boiloff hardware is deferred to the depot, where it sits behind a reactor [16].

## 6.2 Mass budget and launch

Total hydrogen volume is 1,300–1,550 m³. The mission is one flight article, assembled and fueled in a single LEO departure campaign: a heavy-lift launch of the spacecraft followed by three to four commodity-propellant tanker sorties.

| Element | Mass |
|---|---|
| Lander + deployed elements (released at Ceres) | 5.0 t |
| Survey suite + entry capsule (on OTV) | 1.4 t |
| OTV fixed (4 engines, shield, avionics, structure) | 3.5 t |
| Outbound drop tanks (jettisoned at Ceres) | ~10.5 t |
| Return tankage | ~1.5 t |
| $LH_2$, outbound | ~81 t |
| $LH_2$, return (TEI 6.9 + capture 1.4-1.8; incl. boiloff margin Section 6.1) | ~11 t |
| IMLEO (nominal) | ~114 t |
| IMLEO (with 20% system margin) | ~136 t |

*Table 3. Pathfinder mission mass budget.*

## 6.3 Timeline

For calibration, Dawn needed 3.9 years merely to reach its first target with no landing and no return, and a two-spacecraft campaign either launches both elements blind in the same window or adds at least 1.3 years of synodic delay to its second, while facing the same return-entry physics. Every element of the larger architecture, depots, reusable vehicles, propellant production and belt transportation, traces to a capability first demonstrated on this flight. The Pathfinder is Project CERES's first operational realization rather than its precursor.

## 6.4 Concept of operations and timeline (January 2037 window)

| Phase | Duration | Content |
|---|---|---|
| LEO departure campaign | ~3-4 days | Gated perigee burns + perigee-anchored injection; ~5-6% gravity loss |
| Transit out | 370-400 d | Cruise; instrument checkout; dormancy demonstration |
| Ceres arrival (~Jan 2038) | - | Split-burn capture; +1.8% over impulsive |
| Orbital survey | 4-5 mo | Volatile mapping, ice-table sounding, site down-select vs Dawn basemap |
| Surface phase | 4-6 wk | Lander descent, extraction demo, sample loading, ascent, container capture |
| Extended survey & site monitoring | ~20 mo | Return-window phasing converted to program value: seasonal volatile monitoring, candidate-site change detection, comm-relay demonstration for Phase 1 |
| Trans-Earth injection (Jun 2040) | - | Drop tanks jettisoned; TEI 6.9 km/s (survey value) |
| Transit home | 330 d | Cruise; capsule targeting |
| Earth arrival (May 2041) | - | Capsule entry from 12.7 km/s to well beyond 17 (heritage class; Category V handling); OTV lunar-assisted propulsive capture to cislunar (NRHO-class) - mission complete |

*Table 4. Pathfinder Concept of Operations.*

# 7. Beyond the belt

The node's purpose lies beyond Ceres. A vehicle that tanks there begins every journey two-thirds of the way out of the Sun's gravity well, from an orbit in which the outer solar system is, in the only sense that matters, downhill: departure hyperbolic excess to Jupiter is 2.6 km/s from Ceres against 8.8 km/s from Earth, and solar-system escape requires 7.4 km/s of excess against 12.3. Table 5 gives representative refueled departures for a 900 s vehicle from low Ceres orbit. Every entry is an ordinary tank of locally produced propellant; none is a heroic mission, each is a sortie.

| **Destination** | **Departure** | **Flight time** | **ΔV total (km/s)** | **Mass ratio (900 s)** |
|---|---|---|---|---|
| Vesta | Jun 2046 | 2.6 yr | 2.20 | 1.28 |
| Jupiter system | Mar 2049 | 2.5 yr | 5.45 | 1.85 |
| Saturn system | Mar 2046 | 4.9 yr | 6.61 | 2.11 |
| Neptune system | Feb 2043 | 11.0 yr | 11.07 | 3.51 |
| Pallas | Dec 2050 | 3.2 yr | 12.36 | 4.06 |

*Table 5. Refueled departures from low Ceres orbit, 900 s specific impulse, from ephemeris-grade scans. Totals include departure and capture into a loosely bound orbit at the stated periapsis (100,000 km at Jupiter, 70,000 km at Saturn, 30,000 km at Neptune; low orbit at Vesta and Pallas).*

The raw departure ΔV from low Earth orbit is not dramatically worse than from Ceres, because Earth's gravity well confers a strong Oberth advantage; the argument is provenance and posture. Every kilogram burned departing Ceres was melted out of the ground beside the pad, the departing vehicle is a fleet asset on its next sortie rather than a bespoke stack assembled and expended, and the capture propellant that dominates any outer-planet orbiter is delivered to the departure point at local production cost. The deepest change is on time: a Neptune orbiter or Triton lander at mass ratio 3.5 arrives eleven years after departure (Fig. 8), inside one professional career, with a Saturn orbiter in about five years and a Jupiter orbiter in two and a half. Pallas, Ceres' twin in orbit size but inclined 34.9°, calibrates the claim: the node does not make the belt uniformly cheap, and a 12.4 km/s plane-change sortie at mass ratio 4.1 is precisely where high-thrust, high-specific-impulse propulsion is irreplaceable. The node converts the belt, the giant planets and the ice giants from separate mountains into exits on one highway, and whoever builds it first will, for a period that history suggests could be long, define deep-space logistics.

# 8. Economics

A transportation node justifies itself only if it reduces the cost of deep-space access relative to launching every mission independently from Earth. We evaluate it on avoided-cost accounting with every parameter explicit: Phase 1 node hardware at $0.75B delivered as a single 35 t sortie; transfer vehicles at $0.5B amortized over a ten-sortie reactor life; and propellant valued at import parity, which is the 6.3 kg-in-LEO-per-kg-at-Ceres gear ratio times launch cost, or $630 to $6,300 per kilogram of water at the node across a $100–1,000/kg launch-cost range. Local production cost is expected to fall far below import parity but is not yet established by a hardware cost model; the spread is the node's margin.

The skeptic's remaining position is a different architecture rather than a different propulsion choice: if the hub constrains the options, decline the hub. That position deserves a number, because "no hub" is a specific architecture with a specific price, in which every mission carries all of its propellant for every leg from the bottom of Earth's gravity well. A nuclear-thermal vehicle delivering 35 t and returning to cislunar space closes at 220 t in LEO when it refuels at the node, against 556 t carrying its round trip from Earth. With finite-burn losses applied to both, the node cuts LEO launch mass per delivery by 61% before any credit for vehicle reuse, and the launch component of delivered cost falls from \$1,590–15,900/kg to \$630–6,300/kg. Each round-trip sortie that refuels at Ceres avoids 338 t of launch mass; each outbound deep-space departure that tanks there substitutes about 140 t of Earth-shipped hydrogen-equivalent. Beyond Ceres the counterfactual is starker, because without a refueling point at 2.77 AU every outer-system mission reverts to the bespoke pattern of the last sixty years: full injection propellant from Earth, one mission per stack, no fleet, no schedule. The choice is between the hub and that pattern.

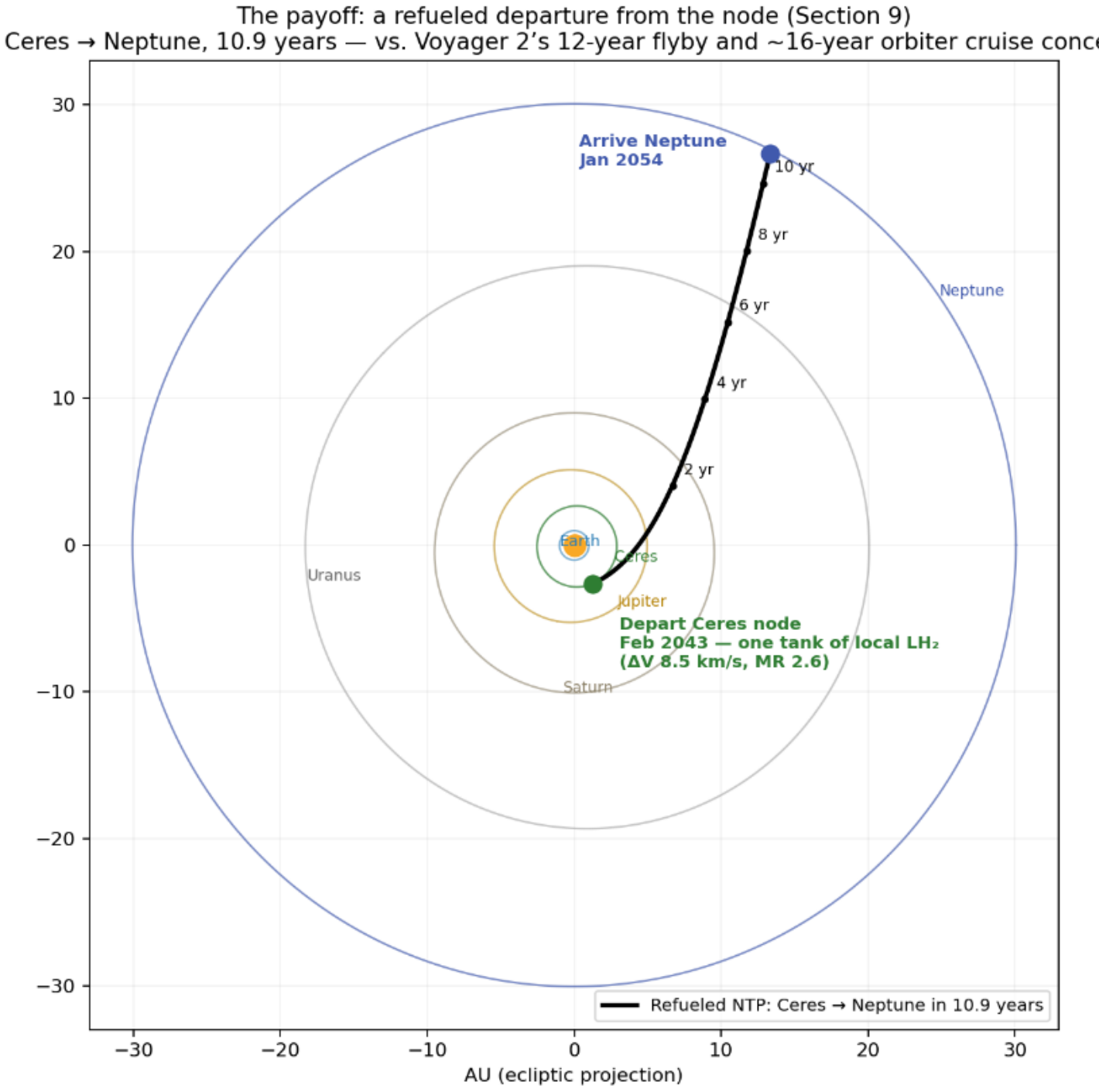

*Figure 8. A refueled departure (Ceres to Neptune), ephemeris-grade solution (depart Feb 2043, 10.9-year transit; rendezvous total 11.1 km/s, mass ratio 3.5 with capture). The trajectory crosses Jupiter's orbit at 1.1 years, Saturn's at 2.6, Uranus's at 6.3, and is itself solar-hyperbolic at 7.5 km/s excess.*

Fig. 9 shows cumulative net value against the Phase 1 investment for three launch-cost regimes under steady traffic of one sortie per window and under a growth profile of one to three sorties per window plus one outer-system tanking sale per two windows from year five. Breakeven arrives at 3.8 years in a \$1,000/kg world, 5.1 years at \$300/kg with growth, and 10.2 years at \$100/kg. The two-regime structure is the point: in expensive-launch worlds the node repays through avoided mass, quickly; in cheap-launch worlds the same cheapness multiplies traffic and the node repays through volume. The investment is hedged against the launch market's central uncertainty rather than exposed to it. Sensitivities across payload class (10–100 t), a steam-only node (Phase 1 alone captures most of the 61%), and node hardware cost (\$0.25–2B) move breakeven near-linearly, and even the pessimistic corner repays within the operating life of a single vehicle fleet. Avoided-cost accounting also deliberately undercounts, since it books no value for missions that cannot fly at all without the node.

The node's market is defined by propellant compatibility, and every propulsion system that expels volatiles is a customer regardless of vendor: nuclear-thermal vehicles on hydrogen, ammonia or steam, chemical stages tanking locally produced hydrolox, and the surface hoppers themselves.

# 9. The Long View

A permanent transportation node at Ceres is technically achievable using propulsion and resource-utilization technologies that are demonstrated or nearing maturity. Such a node would fundamentally change the economics of deep-space exploration. The decision chain runs as follows:

- A robust solar-system logistics network requires an economy of volatiles.
- The volatiles economy requires a transport node.

- The node requires a propulsion class that can hold a schedule against the orbital mechanics quantified in Section 3.
- The chain necessarily begins with a comprehensive Pathfinder campaign to the selected node, Ceres.

The architecture is robotic end to end: the standup case is made without the mass, cost, and schedule burden of life support. But a node that stores hundreds of tonnes of water, manufactures nitrogen and oxygen, and receives scheduled heavy transport has assembled most of what sustained human presence beyond Mars requires. We take no position here on when this happens - only that the architecture keeps the door open at no cost to the baseline.

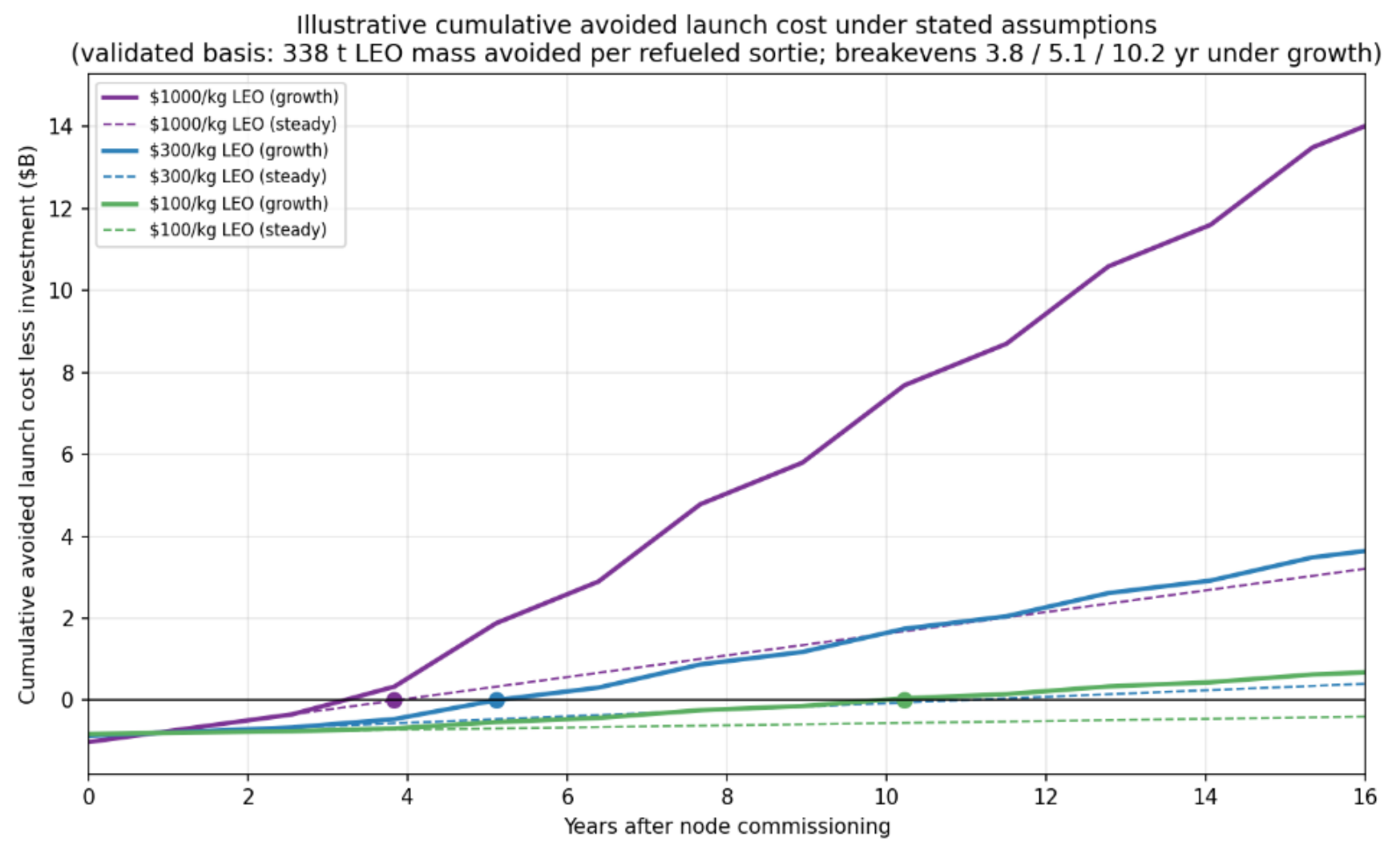

*Figure 9. Cumulative avoided launch cost under stated assumptions (validated basis: 338 t LEO mass avoided per refueled sortie; breakeven 3.8 / 5.1 / 10.2 years at $1,000 / $300 / $100 per kg under the growth profile, which assumes traffic elasticity).*

What this paper rests on is proven; what it proposes is not, and the distinction is key. Rover/NERVA proved the physics: solid-core nuclear reactors at the required performance class, restarted repeatedly, on the ground. Dawn proved the destination: the volatiles are measured, not assumed. What no analysis can prove is a flight-validated NTP stage or a paying manifest, and we claim neither. Project CERES is built for that gap: a sequence of steps small enough to decide individually, each returning standalone value, each generating the evidence on which the next commitment stands.

This paper began with three histories: a missing planet, an ancient ocean world, and six decades in which our reach was constrained less by physics than by our choice of propulsion.

It closes by proposing to open a fourth.  It begins with one reusable spacecraft, nuclear-powered, and a hole in the ice.

# Acknowledgments

We thank our many mentors and colleagues for inspiring in us the desire to reach out, explore, and settle the solar system – and for bequeathing us the ability to find ways to do it effectively and efficiently. We used generative AI tools for editorial assistance, software support, and independent numerical cross-checking. We reviewed all outputs and accept responsibility for the text, analysis, and conclusions.

# Data and Code Availability

All numerical results, figures, and tables in this paper are reproducible from the supplementary package (CERES_Supplementary_v3.2-S1), which constitutes the paper's numerical baseline. The package contains the complete analysis toolchain: ephemeris preparation, the Lambert survey and window scans, gravity-assist census, finite-burn analysis, return-trajectory survey, and the scripts that regenerate every figure and table from public ephemerides and the stated models, with built-in consistency checks. It is archived at Zenodo under DOI 10.5281/zenodo.22307927, license CC BY 4.0; package integrity is verifiable against the SHA-256 checksum published in the deposit record.

# Notes and References